\documentclass[preprints,article,accept,pdftex,moreauthors]{Definitions/mdpi} 

\usepackage{physics}
\firstpage{1} 
\pubvolume{1}
\issuenum{1}
\articlenumber{0}
\pubyear{2025}
\copyrightyear{2025}
\datereceived{ } 
\daterevised{ } 
\dateaccepted{ } 
\datepublished{ } 
\hreflink{https://doi.org/} 

	\Title{Oscillator Algebra in Complex Position-Dependent Mass Systems}
	
	\TitleCitation{Title}
	
	\Author{M.I. Estrada-Delgado $^{1,*,\ddagger}$\orcidA{} and Z. Blanco-Garcia $^{1,*,\ddagger}$\orcidB{} }
	
	\AuthorNames{Firstname Lastname, Firstname Lastname and Firstname Lastname}
	
	\isAPAStyle{%
		\AuthorCitation{Lastname, F., Lastname, F., \& Lastname123456789, F.}
	}{%
		\isChicagoStyle{%
			\AuthorCitation{Lastname, Firstname, Firstname Lastname, and Firstname Lastname.}
		}{
			\AuthorCitation{Lastname, F.; Lastname, F.; Lastname, F.}
		}
	}
	
	\address{%
		$^{1}$ \quad Tecnologico de Monterrey, School of Engineering and Science, Atizapan 52926, Mexico
	}
	
	\corres{Correspondence: iestrada@tec.mx (M.I.E.D);  zblanco@tec.mx (Z.B.G.)}
	
	\secondnote{These authors contributed equally to this work.}
	
	\abstract{This work introduces non-Hermitian position-dependent mass Hamiltonians characterized by complex ladder operators and real, equidistant spectra. By imposing the Heisenberg--Weyl algebraic structure as a constraint, we derive the corresponding potentials, ladder operators, and eigenfunctions. The method provides a systematic procedure for constructing exactly solvable models for arbitrary mass profiles. Specific cases are illustrated for quadratic, cosenoidal, and exponential mass functions.}
	

\begin{document}
		
		\section{Introduction}
		The study of position-dependent mass (PDM) systems has been extensive since 1966, when BenDaniel and Duke, investigating charge carrier behavior in semiconductor heterostructures, proposed a Hermitian Hamiltonian that incorporates the spatial variation of the electron's effective mass across material interfaces \cite{BD1966}. This variation is particularly relevant in systems such as the compositionally graded alloy $\mathrm{Al}_x\mathrm{Ga}_{1 - x}\mathrm{As}
		$, or in abrupt heterojunctions like $\mathrm{InAs}/\mathrm{GaSb}
		$ \cite{VR1983,Bastard1981,Geller1993}. 
		These PDM models find applicability in the design of quantum dots, quantum wells, superlattices, and more general heterostructures \cite{Bayrak2025,Panahi2013,Bastard1981,EN2020,KKS2005,NCA2022}. In such systems, the effective mass profile $m(x)$ often reflects the underlying material composition and can, in principle, be engineered through epitaxial growth techniques and related methods \cite{DW2019}. However, achieving arbitrary spatial profiles remains experimentally challenging. This limitation motivates an alternative theoretical question: given a fixed mass profile, regardless of its intricacy, what forms of the potential $V(x)$ allow the system to exhibit a desired energy spectrum? Addressing this inverse problem opens the door to spectral engineering within constrained material platforms.
		Traditional approaches fix the mass and potential profiles, see for example \cite{Dong2022}, where a particular exponential mass and three different potentials are explored. Other works use supersymmetric quantum mechanics to impose a factorization by using intertwining operators and the superpartners of the systems \cite{Plastino1999,Bagchi2005,Cruz2009,EF2019,Medjenah2023}. Alternative treatments rely on point canonical transformations that map the PDM to a position-independent mass (PIM) Schrödinger equation; once in the regime of PIM, usual exactly solvable techniques are enabled \cite{JYJ2005,GDSGB2023}.
		
		The determination of the spectral profile of PDM systems is a challenging problem in its own right. To date, non-Hermitian, position-dependent, and exactly solvable systems remain largely unexplored, with notable exceptions such as the work of Mustafa and Mazharimousavi, who introduced $\eta$-pseudo-Hermiticity generators for such Hamiltonians \cite{Mustafa2006,MM2008}.

		While the present work focuses on constructing complex ladder operators compatible with such configurations, future extensions could incorporate tools such as supersymmetric quantum mechanics to systematically manipulate the spectrum.
		
		This paper is organized as follows: Sections \ref{sec2} and \ref{sec3} provide a brief review of PDM Hamiltonians and the Heisenberg–Weyl algebra, respectively. In Sec.~\ref{sec4}, we construct a complex first-order ladder operator and the associated non-Hermitian Hamiltonian, present the factorization of both $H$ and $H^\dagger$, describe their commutation relations, and analyze the biorthogonality of the systems. Sec. \ref{sec5} presents illustrative applications involving three distinct mass profiles. Finally, concluding remarks and discussions are offered in Sec. \ref{sec6}.

		\section{Position dependent mass systems}\label{sec2}
		
		One of the most general and Hermitian approaches to quantum PDM systems is the one proposed by von Ross in 1983 \cite{BD1966}:
		
		\begin{equation}
			H_{VR} = \frac{1}{4} \left( m^{\alpha} \hat{p} m^{\beta} \hat{p} m^{\gamma} + m^{\gamma} \hat{p} m^{\beta} \hat{p} m^{\alpha} \right) + V,
			\label{HVR}
		\end{equation}
		
		\noindent
		where the mass, $m(x)$, varies with position. The parameters $\alpha$, $\beta$, and $\gamma$ are real constants that satisfy $\alpha + \beta +\gamma = -1$. Hermiticity is maintained as long as $V(x)$ is real.
		
		\subsection{BenDaniel-Duke Hamiltonians}
		
		It is possible to manipulate the expression in equation \eqref{HVR} so that an effective potential emerges. The resulting expression is:
		
		\begin{linenomath}
			\begin{equation}
				H_{VR}=-\frac{\hbar^{2}}{2m(x)}\frac{d^{2}}{dx^{2}}+\frac{\hbar^{2}m'(x)}{2m^{2}(x)}\frac{d}{dx}+V_{\text{eff}}(x),
				\label{VR2}
			\end{equation}
		\end{linenomath}
		
		\noindent
		where
		\begin{linenomath}
			\begin{equation}
				V_{\text{eff}}(x)=V(x)+\hbar^{2}\left(\frac{(1+\beta)m''(x)}{4m^{2}(x)}-\frac{[m'(x)]^{2}}{2m^{3}(x)}(\alpha^{2}+\alpha\beta+\alpha+\beta+1)\right).
			\end{equation}
		\end{linenomath}
		
		\noindent
		Notice that equation \eqref{VR2} is the one proposed by BenDaniel and Duke in 1966 \cite{BD1966}.
		
		Given the equivalence between the von Ross Hamiltonian and the BDD Hamiltonian, from now on, this work will focus on the BDD Hamiltonian form, keeping in mind that the connection between the BDD and von Ross Hamiltonians, as well as any other Hamiltonian derived from it, is straightforward. Importantly, unlike other treatments that explicitly address the operator ordering ambiguity (see, e.g., \cite{DA2000}), this work avoids the issue altogether by employing an effective potential that already accounts for all ordering-related contributions. This allows us to proceed without fixing specific values of the ordering parameters, maintaining flexibility in the mass profile without complicating the formulation.

		\section{The Heisenberg–Weyl algebra}\label{sec3}
		
		The harmonic oscillator satisfies the well-known Heisenberg-Weyl algebra, where a Hamiltonian, $H$, and two ladder operators, $a^{\pm}$, satisfy the following algebraic relationships:

		\begin{linenomath}
			\begin{equation}
				[H,a^{\pm}]=\pm a^{\pm}, \hspace{1cm} [a^+,a^-]=1.
				\label{HOSCR}
			\end{equation}
		\end{linenomath}
		
		\noindent
		When working with the prototypical example of the harmonic oscillator, this algebraic relationship allows for the determination of the ground state, which is annihilated by $a^-$ and satisfies

		\begin{linenomath}
			\begin{equation}
				a^- \ket{\psi}=0.
			\end{equation}
		\end{linenomath}
		
		\noindent
		Using the previously mentioned commutation relation, all excited states can be generated through the iterative operation of $a^+$.
		
		\section{Complex first order ladder operators on the BDD Hamiltonian}
		\label{sec4}
		
		The search of a general PDM system, specifically a BenDaniel Duke Hamiltonian possessing first-order ladder operators, was already initiated in \cite{EF2019}. However, that work was restricted to Hermitian Hamiltonians and real ladder operators. In this paper, we aim to generalize this study to include non-Hermitian Hamiltonians and complex ladder operators, satisfying a Heisenberg-Weyl algebra. To this end, we begin with the aforementioned Hamiltonian:
		
		\begin{linenomath}
			\begin{equation}
				H=-\frac{\hbar ^2 }{2m(x)}\frac{d^2 }{d x^2} + \frac{\hbar ^2 m'(x) }{2m^2(x)} 
				\frac{d }{d x} + V(x).
			\end{equation}
		\end{linenomath}
		
		\noindent
		Here, $V(x)$ is allowed to be a complex potential. We also need to construct a first-order differential operator, 
		
		\begin{linenomath}
			\begin{equation}
				A^- = \frac{1}{\sqrt{2}}\left(\alpha (x) \frac{d }{dx}+\beta_R(x)+i\beta_I(x)\right),
			\end{equation}
		\end{linenomath}
		
		\noindent
		following the same idea presented in \cite{EF2019}, but allowing for a complex $\beta(x)$ function such that $\beta(x) = \beta_R(x)+i\beta_I(x)$. This operator is meant to act as a complex ladder operator, specifically an annihilator operator $A^-$. For this to hold, a commutation relation, similar to the one presented in equation \eqref{HOSCR}, must be satisfied,  
		
		\begin{linenomath}
			\begin{equation}
				[H,A^-] = -\Delta E A^-, \hskip2ex \Delta E \in \mathbb{R}.
			\end{equation}
		\end{linenomath}
		
		\noindent
		After imposing this commutativity  condition, the following equations arise:
		
		\begin{linenomath}
			\begin{eqnarray}
				&&\alpha'(x)+\frac{m'(x)}{2 m(x)}\alpha(x) = 0 \label{expr3},\\ 
				&&\beta_R'(x) = \frac{\alpha (x) [m'(x)]^2}{m^2(x)}+\frac{m'(x) \alpha '(x)-\alpha (x) m''(x)}{2 m(x)}+\frac{\text{$\Delta $E} m(x) \alpha (x)}{\hbar ^2}-\frac{\alpha ''(x)}{2} \label{betaRp},\\
				&&\beta_I'(x) = 0 \label{betaIp},
			\end{eqnarray}
		\end{linenomath}

		\begin{equation}
			\hspace{-2.2cm}V'(x) = \frac{1}{2 m^2(x) \alpha (x)} \left[ 2\Delta E m^2(x) \left[\beta_R(x) + i \beta_I(x) \right] + \hbar^2 m'(x) \left[\beta'_R(x) + i \beta'_I(x) \right] + \hbar^2 m(x) \left[-\beta''_R(x) - i \beta''_I(x) \right]\right].
			\label{pVoltage}
		\end{equation}

		\noindent
		We can solve equation \eqref{expr3} to find $\alpha(x)$ as a function of the mass as follows:
		
		\begin{linenomath}
			\begin{eqnarray}
				\alpha (x) = \frac{a}{\sqrt{m(x)}}.
			\end{eqnarray}
		\end{linenomath}
		
		\noindent
		From \eqref{betaIp} it is concluded that $\beta_I = \lambda $, where $\lambda$ is an integration constant. Subsequently, the substitution of $\alpha(x)$ as a function of $m(x)$ leads to a general expression for $\beta_R(x)$ as a function of the mass.
		
		\begin{linenomath}
			\begin{eqnarray}
				\beta_R(x) = \frac{a}{2} \left([m(x)]^{-1/2}\right)' + \frac{a \Delta E}{\hbar^2} F(x), \hskip2ex  F(x) = \int \sqrt{m(x)} dx.
			\end{eqnarray}
		\end{linenomath}
		
		\noindent
		The previous equation resembles greatly the $\beta$ function presented in \cite{EF2019} when the ladder operator $A^-$ was real. 
		
		Incorporating all of the above expressions, the potential is determined as

		\begin{linenomath}
			\begin{equation}
				V(x) = V_R(x) + i V_I(x),
			\end{equation}
		\end{linenomath}
		where the real and imaginary parts are given by:
		\begin{linenomath}
			\begin{eqnarray}
				&&V_R(x) = \frac{1}{2} \left(\frac{\Delta E}{\hbar}\right)^2 F^2(x) -\frac{\hbar^2}{8} \left(\frac{7[m'(x)]^2}{4m^3(x)}  - \frac{m''(x)}{m^2(x)}\right),\\
				&&V_I(x) = \frac{\Delta E}{a} \lambda F(x).
			\end{eqnarray}
		\end{linenomath}
		
		\noindent
		It is worth noting that the presence of a complex component in the ladder operator, introduced through a constant $\lambda$, results in a nontrivial modification of the potential.
		
		Looking for the ground state, at least a mathematical one, we should look for a function that is annihilated by $A^-$. Such an eigenfunction is given as follows:
		
		\begin{linenomath}
			\begin{equation}
				\psi_0 = c_0 [m(x)]^{1/4} \exp \left(-\frac{\Delta E }{2 \hbar ^2} F^2(x)\right) \left(\cos (\lambda F(x))-i \sin (\lambda F(x))\right).
			\end{equation}
		\end{linenomath}
		
		\noindent
		Let us remark that the current Hamiltonian $H$ is not Hermitian; therefore, the operator $(A^-)^\dagger$ acts as a creation operator for $H^+$, but not, as desired, for $H$. Thus, we need to begin the search for the corresponding creation operator.
		
		We propose the existence of a creation operator $A^+$ that is akin to the dagger of $A^-$ but not exactly the same:
		
		\begin{linenomath}
			\begin{equation}
				A^+ =\frac{1}{\sqrt{2}}\left( -\alpha(x)\frac{d}{dx} - \alpha'(x) + \gamma(x)\right),
			\end{equation}
		\end{linenomath}
		
		\noindent
		where $\gamma(x) = \gamma_R(x) + i\gamma_I(x)$. With this proposition of $A^+$ and after the imposition of the corresponding commutation relation $[H,A^+]=\Delta E A^+$, we end up with the following creation operator:
		
		\begin{linenomath}
			\begin{equation}
				A^+ =\frac{1}{\sqrt{2}}\left( -\alpha(x)\frac{d}{dx} - \alpha'(x) + \beta_R(x) + i \beta_I(x)\right).
			\end{equation}
		\end{linenomath}
		
		It is worth noticing that the only difference with respect to the adjoint of $A^-$ is a sign in the complex part.
		
		Since the creation operator has been constructed, we are now prepared to build the possible excited states by repeatedly applying $A^+$ to the ground state; thus 
		\begin{linenomath}
			\begin{equation}
				\psi_n =c_n (A^+)^n \psi_0.
			\end{equation}
		\end{linenomath}

		The eigenvalues associated with these states, and possibly the physical energies once the boundary conditions are satisfied, are given as follows:
		
		\begin{linenomath}
			\begin{equation}
				E_n = \left(n+\frac{1}{2}\right) \Delta E + \frac{1}{2}\lambda^2 \hbar^2, \hskip1ex n=0,1,2,\dots
			\end{equation}
		\end{linenomath}
		
		Finally, the commutator between the two ladder operators $A^+$ and $A^-$ can be readily calculated as 
		
		\begin{linenomath}
			\begin{equation}
				\left[A^-, A^+\right] = \frac{a^2\Delta E}{\hbar^2}.
			\end{equation}
		\end{linenomath}
		
		This commutation relation corresponds to the zero-degree polynomial Heisenberg algebra, where the commutator reduces to a constant.
		
		\subsection{Factorization}
		
		Since $H^{\dagger} \neq H$, the system is no longer Hermitian. This implies that orthogonality is no longer guaranteed.
		
		Let us point out that the following commutation relation is satisfied:
		
		\begin{linenomath}
			\begin{equation}
				[H,A^{\pm}]=\pm \Delta E A^{\pm},
			\end{equation}
		\end{linenomath}
		
		\noindent
		after defining the adjoint ladder operators as:
		\begin{linenomath}
			\begin{equation}
				(A^{\pm})\dagger=B^\mp,
			\end{equation}
		\end{linenomath}
		
		\noindent
		it follows straightforwardly that $B^{\pm}$ are indeed the ladder operators of $H^{\dagger}$:
		
		\begin{linenomath}
			\begin{eqnarray}
				\left[H,A^{\pm}\right]^{\dagger} &=& \pm \Delta E (A^{\pm})^{\dagger},\nonumber\\
				\left[H^{\dagger},B^{\mp}\right]&=&\mp \Delta E B^{\mp},\nonumber\\
				\left[H^{\dagger},B^{\pm}\right]&=& \pm \Delta E B^{\pm}.
			\end{eqnarray}
		\end{linenomath}
		
		At this point, we have two Hamiltonians with their respective ladder operators and, as is expected, both can be factorized in terms of their respective ladder operators as follows:
		\begin{linenomath}
			\begin{eqnarray}
				H&=A^{+}A^{-}+E_0,\\
				H^{\dagger}&=B^{+}B^{-}+E_0.
			\end{eqnarray}
		\end{linenomath}

		The eigenstates of $H^{\dagger}$ are constructed analogously:
		\begin{linenomath}
			\begin{equation}
				B^{-}\phi_0=0, \hspace{1cm}\phi_n=\overline{c_n} (B^{+})^n\phi_0\hspace{1cm}n\in \mathbb{N},
			\end{equation}
		\end{linenomath}
		where $\phi_n$ are the eigenfunctions of $H^{\dagger}$. After simplification, it is easy to verify that:
		\begin{linenomath}
			\begin{equation}
				\phi_n=\psi_n^{*},
			\end{equation}
		\end{linenomath}
		and that both sets share the same real eigenvalues. 
		
		\subsection{Biorthogonality}
		
		From the mathematical relations established above, it follows that:
		\begin{linenomath}
			\begin{eqnarray}
				\langle \phi_m | H \psi_n \rangle &=& E_n \langle \phi_m | \psi_n \rangle, \nonumber\\
				\langle H^\dagger \phi_m | \psi_n \rangle &=& E_m \langle \phi_m | \psi_n \rangle, \nonumber\\
				(E_n - E_m) \langle \phi_m | \psi_n \rangle &=& 0.
			\end{eqnarray}
		\end{linenomath}
		
		Hence, the inner product $\langle \phi_m | \psi_n \rangle$ must vanish for all $m\neq n$. Nevertheless, according to \cite{CM2007,M2011}, the biorthogonal inner product — distinct from the conventional one — is defined as:
		
		\begin{linenomath}
			\begin{equation}
				\langle \phi_m | \psi_n \rangle=\int_{\Omega}\phi(x)\psi(x)dx,
			\end{equation}
		\end{linenomath}
		but, since we have already stated that $\phi_n=\psi_n^*$, we surprisingly recover a more familiar relation:
		
		\begin{linenomath}
			\begin{equation}
				\langle \phi_m | \psi_n \rangle=\int_{\Omega}\psi_m^*(x)\psi_n(x)dx=\delta_{mn},
			\end{equation}
		\end{linenomath}
		
		It is important to note that we are restricted to an arbitrary, but suitable, choice of a non-negative $m(x)$, with the requirement that this choice, together with appropriate boundary conditions, generates well-defined and normalizable eigenfunctions as described above.

		\section{Implementation }
		\label{sec5}

		In this section, we illustrate the theoretical framework by applying it to specific examples, thereby highlighting its practical implications. To this end, we showcase the following mass profiles:
		
		\begin{linenomath}
			\begin{align}
				m(x) &= m_0(\sigma +\delta x^2), \\
				m(x) &= m_0 + \cos(x), \label{massprofilescos}\\
				m(x) &= \frac{m_0 \alpha}{1 - e^{-\alpha}} e^{-\alpha x}.
				\label{massprofilesE}
			\end{align}
		\end{linenomath}
		
		The constants in each expression are chosen solely for illustrative purposes to highlight simpler cases. Explicit expressions will be shown only for the first example, as their length increases with more complicated mass profiles.

		\subsection{Quadratic mass profile}
		
		We consider a mass profile similar to that presented in \cite{KKS2005}. In this example, we use $m(x) = 1 + x^2$ and $\lambda = \frac{1}{5}$. Given this mass profile, the following functions are calculated.

		\begin{linenomath}
			\begin{align}
				\alpha(x) &= \frac{1}{\sqrt{x^2 + 1}}, \\
				\beta_R(x) &= \frac{1}{2} \left( \frac{(x^2 + 2)\, x^3}{(x^2 + 1)^{3/2}} + \sinh^{-1}(x) \right), \\
				\beta_I(x) &= \frac{1}{5}, \\
				V(x) &= \frac{1}{8} \left( \sqrt{x^2 + 1}\, x + \sinh^{-1}(x) \right)^2
				+ \frac{1}{8} \left( \frac{2}{(x^2 + 1)^2} - \frac{7x^2}{(x^2 + 1)^3} \right) \notag \\
				&\quad + \frac{1}{10} i \left( \sqrt{x^2 + 1}\, x + \sinh^{-1}(x) \right), \\
				\psi_0(x) &= 0.751126\, e^{-\frac{1}{8} \left( \sqrt{x^2 + 1}\, x + \sinh^{-1}(x) \right)^2}
				(x^2 + 1)^{1/4} \notag \\
				&\quad \times \left[
				\cos\left( \frac{1}{10} \left( \sqrt{x^2 + 1}\, x + \sinh^{-1}(x) \right) \right)
				- i \sin\left( \frac{1}{10} \left( \sqrt{x^2 + 1}\, x + \sinh^{-1}(x) \right) \right)
				\right].
			\end{align}
		\end{linenomath}
		
\begin{figure}[h!]
			\centering
			\includegraphics[width=1\linewidth]{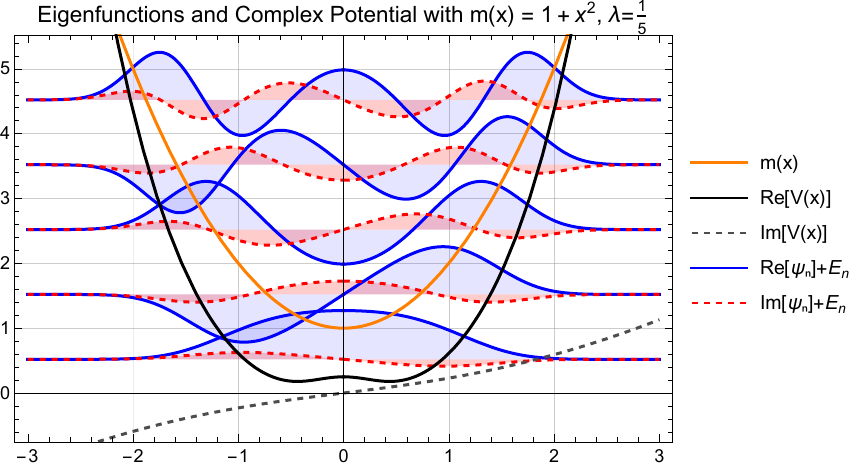}
			\caption{The figure shows the quadratic mass profile $m_0(\sigma +\delta x^2)$ (yellow), the real and imaginary parts of the potential (solid and dashed black lines, respectively), and the first eigenfunctions, with their real and imaginary parts shown in blue and red, respectively.}
			\label{cosprofile}
		\end{figure}

		As mentioned previously, the corresponding excited states can be found by iteratively applying the creation operator. The mass profile, potential, ground, and first excited states are visualized in figure \ref{cosprofile}.
		
\begin{figure}[h!]
			\centering
			\includegraphics[width=0.49\linewidth]{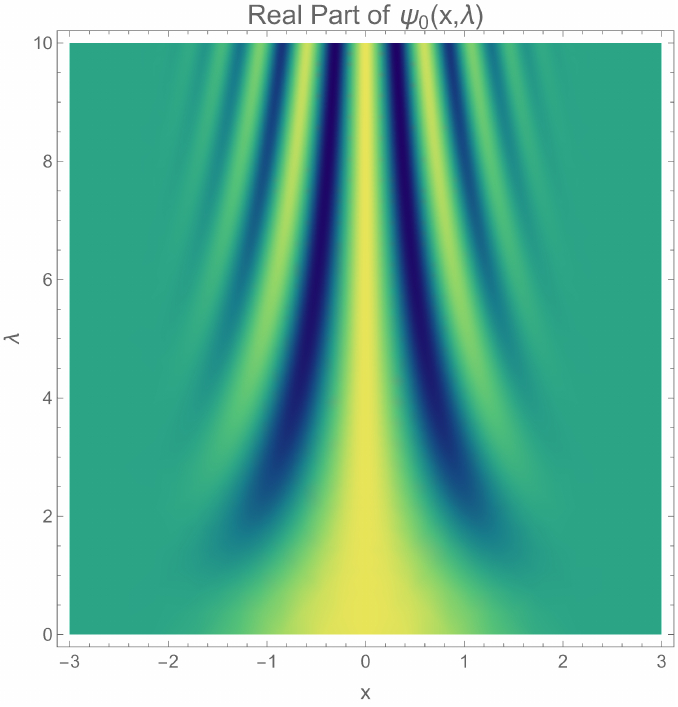}%
			\includegraphics[width=0.49\linewidth]{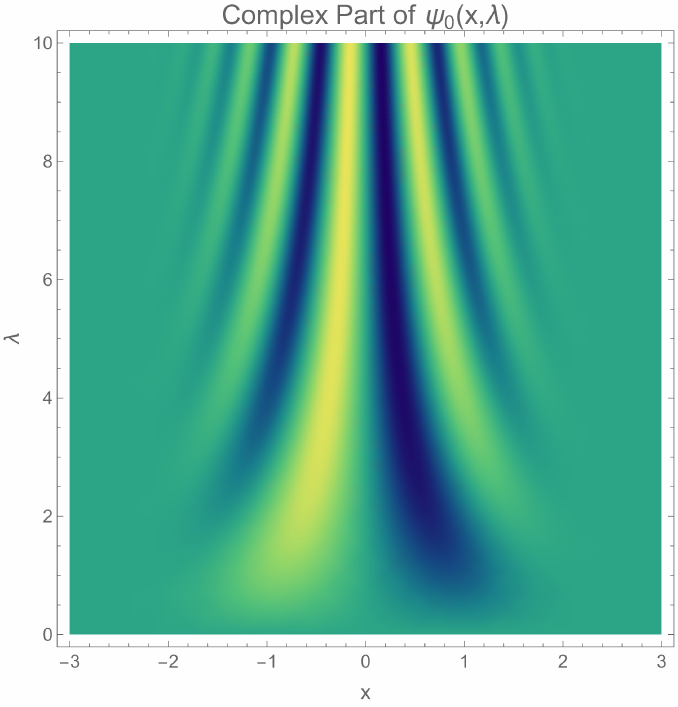}
			\caption{The figure shows the real (left) and imaginary (right) parts of the ground state for the quadratic mass profile as the parameter $\lambda$ increases.}
			\label{prixlambda1plusxpow2}
		\end{figure}

		Notice that the parameter $\lambda$ controls the imaginary parts of the potential and, consequently, the eigenfunctions. The higher the $\lambda$ parameter is, the higher the number of oscillations that we can encounter, even in the ground state. Next, the real and imaginary parts of $\psi_0(x,\lambda)$ are shown in Figure \ref{prixlambda1plusxpow2}.
		
		\subsection{Cosenoidal mass profile}
		
\begin{figure}[h!]
			\centering
			\includegraphics[width=1\linewidth]{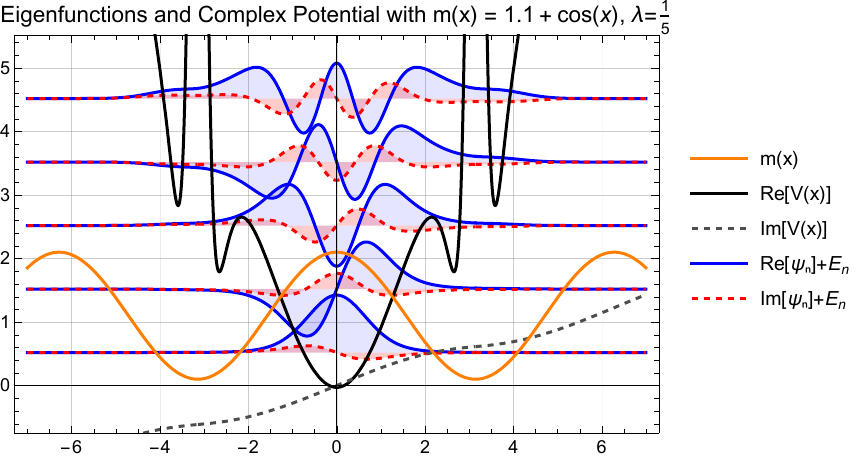}
			\caption{The figure shows the cosenoidal mass profile $m_0 + \cos(x)$ (yellow), the real and imaginary parts of the potential (solid and dashed black lines, respectively), and the first eigenfunctions, with their real and imaginary parts shown in blue and red, respectively.}
			\label{1.1plusCos(x)_p1}
		\end{figure}

		Let us consider the cosenoidal mass profile \eqref{massprofilescos}. After implementing it,  we again obtain a complex potential, and the corresponding eigenstates. See Figure \ref{1.1plusCos(x)_p1}.
		
\begin{figure}[h!]
			\centering
			\includegraphics[width=0.49\linewidth]{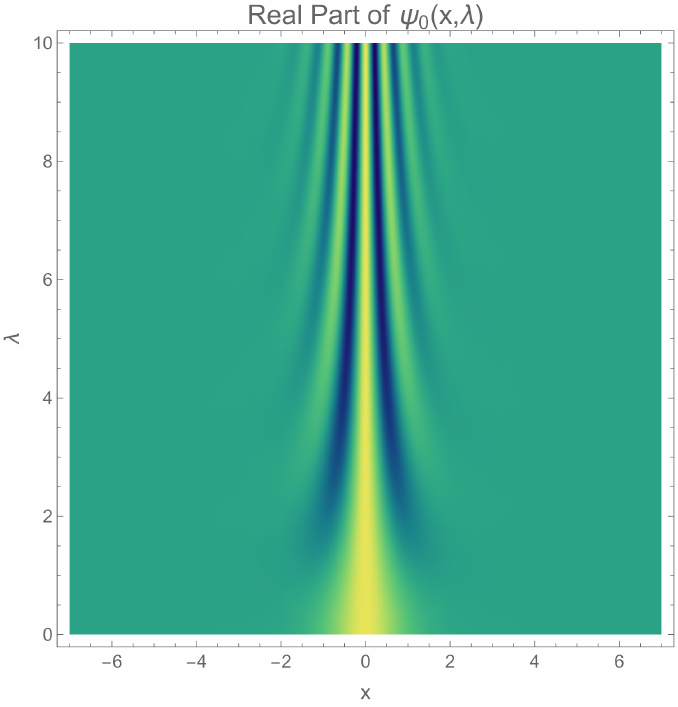}%
			\includegraphics[width=0.49\linewidth]{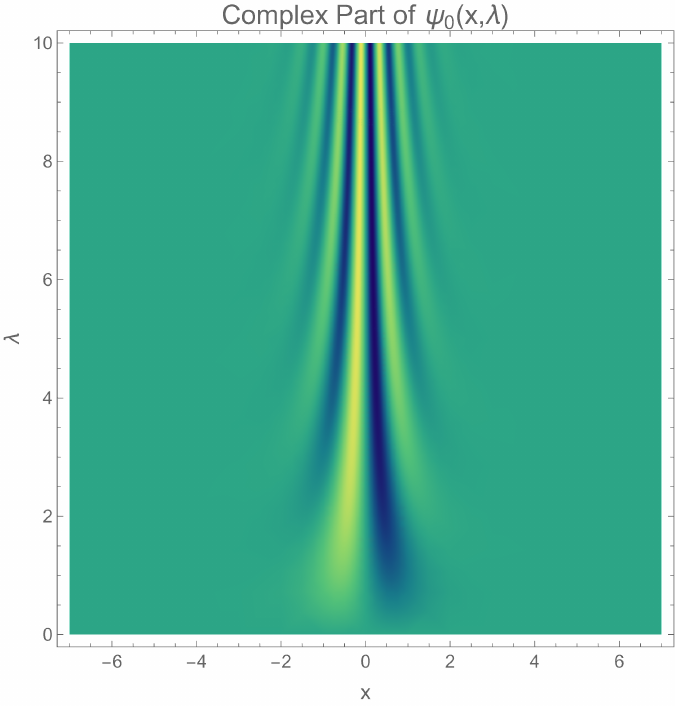}
			\caption{The figure shows the real (left) and imaginary (right) parts of the ground state for the cosenoidal mass profile as the parameter $\lambda$ increases.}
			\label{1.1plusCos(x)}
		\end{figure}

		In the cosenoidal mass profile, the effect of the $\lambda$ parameter is once again evident. As $\lambda$ increases, the number of oscillations increases, even in the ground state. In Figure \ref{1.1plusCos(x)} the real and imaginary components of the ground state $\psi_0(x,\lambda)$ are displayed. 
		
		\subsection{Exponential mass profile}
		
\begin{figure}[h!]
			\centering
			\includegraphics[width=1\linewidth]{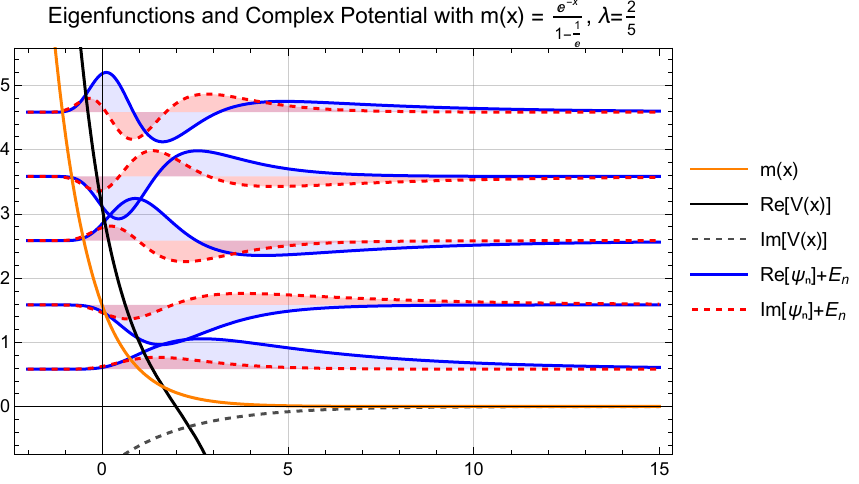}
			\caption{The figure shows the exponential mass profile (yellow), the real and imaginary parts of the potential (solid and dashed black lines, respectively), and the first eigenfunctions, with their real and imaginary parts shown in blue and red, respectively.}
			\label{1((1minusEpow(minus1))xEpowx)_p1}
		\end{figure}

		The exponential mass profile \ref{massprofilesE} illustrated in Figure~\ref{1((1minusEpow(minus1))xEpowx)_p1} was previously analyzed in \cite{Dong2022}. Applying this profile in our framework yields a complex potential, as well as the ground and first excited states of the system.
		
\begin{figure}[h!]
			\centering
			\includegraphics[width=0.49\linewidth]{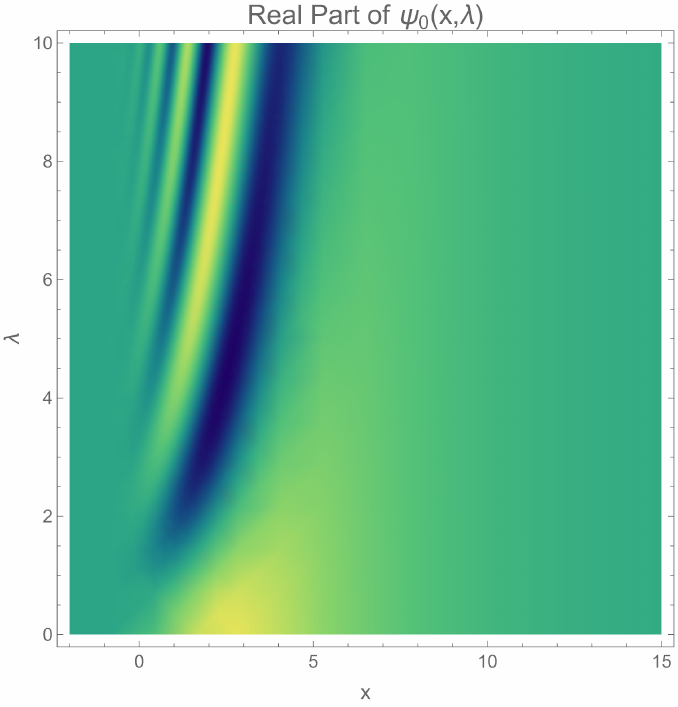}%
			\includegraphics[width=0.49\linewidth]{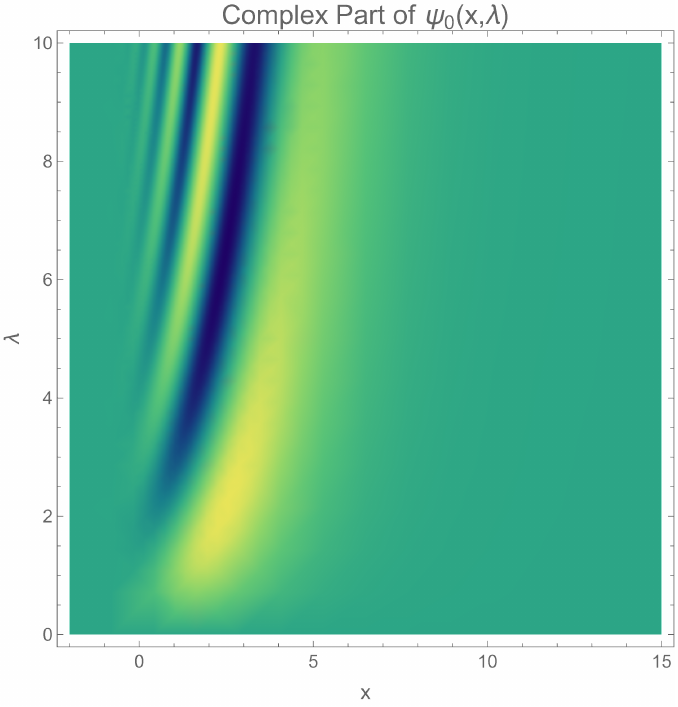}
			\caption{The figure shows the real (left) and imaginary (right) parts of the ground state for the exponential mass profile as the parameter $\lambda$ increases.}
			\label{1((1minusEpow(minus1))xEpowx)}
		\end{figure}

		Figure~\ref{1((1minusEpow(minus1))xEpowx)} illustrates the effect of the $\lambda$ parameter on the ground state, modulating the imaginary part of the potential and its corresponding eigenfunction. As $\lambda$ increases, the number of oscillations in the ground state also increases.

		\section{Discussion}
		\label{sec6}
		
		Non-Hermitian position-dependent mass Hamiltonians with complex ladder operators and real, equidistant spectra have been presented. A first-order differential operator satisfying the Heisenberg–Weyl algebra was constructed based on the BenDaniel–Duke Hamiltonian. To illustrate the implementation, three distinct mass profiles were examined: quadratic, cosenoidal, and exponential. It was found that the parameter $\lambda$, which introduces a displacement in the energy spectrum and governs the imaginary part of the Hamiltonians, significantly affects the behavior of the eigenfunctions. In particular, as $\lambda$ increases, the number of oscillations rises, and the positions of the nodes shift, even in the ground state. Notably, the ground state may exhibit oscillatory behavior solely due to variations in the $\lambda$ parameter.
		
		Furthermore, since the Hamiltonians are non-Hermitian, the eigenstates are not orthogonal in the conventional sense. However, we demonstrated that the eigenstates of $H$ and $H^\dagger$ form a biorthogonal system, preserving a generalized version of the superposition principle and supporting a well-defined ladder operator formalism.
		
		This work focused on the construction of complex ladder operators satisfying the Heisenberg–Weyl algebra, leading to the determination of the complex potential, eigenfunctions, and energy spectrum for an arbitrary mass profile. In future work, we aim to incorporate supersymmetric quantum mechanics as a tool for spectral manipulation.
		
		
		
		\vspace{6pt} 
		
		\authorcontributions{All authors contribute evenly in every section of this work. All authors have read and agreed to the published version of the manuscript.}
		
		\funding{This research received no external funding.}
		
		\institutionalreview{Not applicable.}
		
		\dataavailability{Not applicable.} 
		
		\acknowledgments{Authors acknowledge the economic support to publish this article to the School of Engineering and Science from Tecnologico de Monterrey.}
		
		\conflictsofinterest{The authors declare no conflict of interest.} 
		
		
		
		\abbreviations{Abbreviations}{
			The following abbreviations are used in this manuscript:
			\\
			
			\noindent 
			\begin{tabular}{@{}ll}
				PDM & Position dependent mass\\
				PIM & Position independent mass\\
				BDD & BenDaniel-Duke\\
			\end{tabular}
		}
		
		
		
		\reftitle{References}
		
		
		
		
		\isAPAandChicago{}{%
			
		}
		
		%
		
		
	\end{document}